\documentclass[aps,onecolumn,10pt]{revtex4}
\usepackage{amsmath}
\usepackage{amssymb}
\usepackage{hyperref}
\hypersetup{colorlinks=true,linkcolor=red,citecolor=green,urlcolor=blue}
\numberwithin{equation}{section}
\numberwithin{equation}{section}

\begin{document}
\allowdisplaybreaks
\setcounter{equation}{0}
%\baselineskip=0.9cm
 
%\title{Time Advance in PT-Symmetric Quantum Mechanics and Negative Time Delay}
\title{Time Advance and Probability Conservation in PT-Symmetric Quantum Mechanics}

\author{Philip D. Mannheim}
\affiliation{Department of Physics, University of Connecticut, Storrs, CT 06269, USA \\
philip.mannheim@uconn.edu\\ }

\date{May 13 2025}

\begin{abstract}
When excited states decay the time evolution operator $U(t)=e^{-iHt}$ does not obey $U^{\dagger}(t)U(t)=I$. Nonetheless, probability conservation  is not lost if one includes both excitation and decay, though it takes a different form.  Specifically, if the eigenspectrum of a Hamiltonian is complete, then due to $CPT$ symmetry, a symmetry that holds for all physical systems, there must exist an operator $V$ that effects $VHV^{-1}=H^{\dagger}$, so that $V^{-1}U^{\dagger}(t)VU(t)=I$. In consequence, the time delay associated with  decay must be accompanied by an equal and opposite  time advance for excitation. Thus when a photon excites an atom the spontaneous emission of a photon from the excited state must occur without any decay time delay at all. An effect of this form together with an associated negative time delay appear to have recently been reported by Sinclair et. al., PRX Quantum \textbf{3},  010314 (2022) and Angulo et. al., arXiv:2409.03680 [quant-ph].

\end{abstract}

\maketitle

\section{Time delay and time advance}
 \label{S1}

In conventional nonrelativistic Hermitian quantum mechanical scattering by  a potential well a wave with an energy above the well undergoes a time delay  due to being held for some time by the well as it traverses it. This time delay can be understood as being due to an energy-dependent phase shift of the form $\tan \delta=\Gamma/(E_0-E)$, with $\delta =\pi/2$ when the energy $E$ is at resonance $E=E_0$. With a phase shift of this form the scattering amplitude behaves as 
\begin{align}
f(E)\sim e^{i\delta}\sin\delta= \frac{\Gamma}{E_0-i\Gamma-E},
\label{1.1}
\end{align}
and the propagator has the standard Breit-Wigner form
\begin{align}
D_{\rm BW}(E)=\frac{1}{E-E_0+i\Gamma}
=\frac{E-E_0-i\Gamma}{(E-E_0)^2+\Gamma^2},
\label{1.2}
\end{align}
with a pole at $E=E_0-i\Gamma$. To determine the associated time dependence we Fourier transform and obtain
\begin{align}
D_{\rm BW}(t)=\frac{1}{2\pi}\int _{-\infty}^{\infty}dE e^{-iEt}D_{\rm BW}(E)=\frac{1}{2\pi }\int _{-\infty}^{\infty}dE \frac{e^{-iEt}}{E-E_0+i\Gamma}.
\label{1.3}
\end{align}
With the pole being in the lower half of the complex $E$ plane, and with $e^{-iEt}$ vanishing on the lower-half plane circle at infinity when $t>0$ while vanishing on the upper-half plane circle at infinity when $t<0$, contour integration gives 
\begin{align}
\frac{1}{2\pi }\int _{-\infty}^{\infty}dE \frac{e^{-iEt}}{E-E_0+i\Gamma}=-i\theta(t)e^{-iE_0t-\Gamma t}.
\label{1.4}
\end{align}
Thus  $f(t)$ and $D_{\rm BW}(t)$ only possess a decaying mode, with a time delay $\Delta T(E)=\hbar d\delta/dE$ \cite{Wigner1955} of the form \cite{footnote1}
\begin{align}
\Delta T(E)=\hbar\frac{d\delta}{dE}=\frac{\hbar\Gamma}{[(E-E_0)^2+\Gamma^2]},\qquad \Delta T(E_0)=\frac{\hbar}{\Gamma}.
\label{1.5}
\end{align}

A situation in which a time delay of this form could occur is in the decay of an excited state of an atom, a state that  has an energy $E_0$ above the ground state. Specifically, due to quantum-field-theoretic  electromagnetic fluctuations in the virtual photon cloud that accompanies each atomic electron, the excited state energy level acquires a natural line width $\Gamma$. The excited state then decays down to the ground state with a time delay of order $\hbar/\Gamma$ as virtual photons go on shell and carry off energy $E_0-i\Gamma$. The typical time scale for this decay is $10^{-9}$ seconds if $\Gamma$ is in the $10^{-6}$ eV region. In contrast,  for the initial excitation of the atom from its ground state to the excited state the time scale is of order $\hbar/E_0$, which is  typically of order  the much shorter $10^{-15}$ seconds if $E_0$ is in the eV region. 

However, in and of themselves decays do not conserve probability, with the above analysis thus not being complete. Probability conservation is restored if one includes the decay products, since as the population of the excited level decreases that of its decay products increases by an equal amount. A time delay $\Gamma$ of an $e^{-\Gamma t}$ decay is thus accompanied by the time advance $-\Gamma$ of an $e^{\Gamma t}$ growth. Equally, growth (viz. excitation) can precede the decay.

To get such a possible time advance and see how to apply it to atomic excitations  we need to be able to generate an $E=E_0+i\Gamma$ term. While this cannot be achieved in Hermitian quantum mechanics it can be achieved if the requirement of Hermiticity is extended to the more general requirement of antilinear symmetry (a requirement that includes Hermiticity as a special case), since with antilinear operators serving to complex conjugate numbers the energy $E_0-i\Gamma$ can be brought to an $E_0+i\Gamma$ form by an antilinear transformation. Such an antilinearity requirement is actually the most general requirement that one can put on a quantum theory \cite{Mannheim2018a} and still be fully compatible with quantum mechanics, with quantum mechanics  being richer than a restriction to Hermiticity. While a particular antilinear symmetry, namely $PT$ symmetry   ($P$ is parity, $T$ is antilinear time reversal) was the first antilinear symmetry to be studied in this context \cite{Bender1998} (to thus give the field its ``$PT$ symmetry" name), it was subsequently found that $CPT$ symmetry ($C$ is charge conjugation) was the most general allowable antilinear symmetry \cite{Mannheim2018a}, provided only that one imposed complex Lorentz invariance (viz. invariance under the proper, orthochronous Lorentz group) and probability conservation. Thus imposing $CPT$ symmetry is not discretionary, and any  experiment that one can perform must be  $CPT$ symmetric.  In a $C$-conserving process $CPT$ symmetry  then defaults to the $PT$ symmetry that will be of interest to us here. The existence of  a possible time advance in a theory with an antilinear symmetry theory was noted in \cite{Mannheim2018a}. We thus now discuss how time advance comes about in a theory with antilinear symmetry, and then discuss some experimental consequences.

\section{$PT$ Symmetry and $CPT$ symmetry}
\label{S2}

Through a whole host of papers, such as  \cite{Bender1998,Bender1999,Bender2002,Bender2007,Makris2008,Bender2008a,Bender2008b,Guo2009,Bender2010,Special2012,Theme2013,ElGanainy2018,Bender2018,Mannheim2018a,Fring2021}, together with well over 12,000 published papers to date, antilinear symmetry has been found to be  intrinsic to fundamental  theory. These studies have been triggered by the realization that while Hermiticity of a Hamiltonian requires real eigenvalues, there is no converse theorem that says that if a Hamiltonian is not Hermitian then at least one of  its eigenvalues must be complex. Hermiticity is only sufficient to guarantee real eigenvalues but not necessary.   A necessary condition has been identified in the literature, namely that a Hamiltonian have an antilinear symmetry. This was shown first for finite-dimensional quantum systems in \cite{Bender2010}, and then for infinite-dimensional quantum field theory in \cite{Mannheim2018a}.  If we introduce some general antilinear operator $A$ that commutes with $H$ and consider a Hamiltonian  with eigenvalues $E$ and eigenfunctions $e^{-iEt}\vert \phi\rangle$ that obey $H\vert \phi \rangle=E\vert \phi\rangle$ we obtain
\begin{align}
HA\vert \phi\rangle=AH\vert \phi\rangle=AE\vert \phi\rangle=E^*A\vert \phi\rangle.
\label{2.1}
\end{align}
Thus for every eigenvalue $E$ with eigenvector $\vert \phi\rangle$ there is an eigenvalue $E^*$ with eigenvector $A\vert \phi\rangle$. Thus as first noted by Wigner in his study of time reversal invariance \cite{Wigner1960}, antilinear symmetry implies that energy eigenvalues are either real or in complex conjugate pairs. In  \cite{Mannheim2018a} it was shown that if the eigenspectum of $H$ is complete then antilinear symmetry is a necessary condition for real eigenvalues, with the necessary and sufficient condition being that $AH=HA$ and that every eigenstate of $H$  also be an eigenstate of $A$.

Antilinear symmetry is also the necessary condition for probability conservation, with  Hermiticity only being sufficient but not necessary. Specifically, with $H$, Hermitian or not,  always being the time evolution generator according to $U(t)=e^{-iHt}$, for a general not necessarily Hermitian $H$  we obtain 
\begin{align}
\langle \psi(t)\vert \psi(t)\rangle=\langle \psi(t=0)\vert U^{\dagger}(t)U(t)\vert \psi(t)=\langle \psi(t=0)\vert e^{+iH^{\dagger}t-iHt}\vert \psi(t=0)\rangle.
\label{2.2}
\end{align}
Thus the standard Dirac norm $\langle \psi(t)\vert \psi(t)\rangle$ is not preserved in time if $H$ is not Hermitian, with Hermiticity only being able to secure probability conservation if the norm is the standard Dirac norm. However, that does not mean that there is no other appropriate inner product if $H$ is not Hermitian, it only means that any such inner product cannot be the standard Dirac one. Instead, we introduce a time-independent  operator $V$ and evaluate
\begin{align}
i\frac{\partial}{\partial t}\langle \psi(t)\vert V\vert \psi(t)\rangle=\langle \psi(t)\vert (VH-H^{\dagger}V)\vert \psi(t)\rangle.
\label{2.3}
\end{align}
Thus $\langle \psi(t)\vert V\vert \psi(t)\rangle$ will be time independent
 if there exists a $V$ that obeys 
\begin{align}
VH=H^{\dagger}V,\quad VHV^{-1}=H^{\dagger},
\label{2.4}
\end{align}
with the second condition requiring that $V$ be invertible, something we take to be the case here.
In \cite{Mannheim2013} and \cite{Mannheim2018a} it was shown that this condition is both necessary and sufficient for the time independence of inner products if the eigenspectrum of $H$ is complete. The $VHV^{-1}=H^{\dagger}$ condition is  known as pseudo-Hermiticity, and with $V^{-1}e^{iH^{\dagger}t}V=e^{iHt}$, the $V^{-1}U^{\dagger}(t)VU(t)=I$ condition is known as pseudounitarity. Pseudounitarity thus generalizes ordinary $U^{\dagger}(t)U(t)=I$ unitarity to the non-Hermitian case, while the use of complex Lorentz invariance enables us to generalize the $CPT$ theorem to the non-Hermitian case as well \cite{Mannheim2018a}.

The $VHV^{-1}=H^{\dagger}$ condition entails that the relation between $H$ and $H^{\dagger}$ is isospectral. Thus every eigenvalue of $H$ is an eigenvalue of $H^{\dagger}$. Consequently the eigenvalues of $H$ are  either real or in complex conjugate pairs. But this is the antilinear symmetry condition. Thus for any $H$ whose eigenspectrum is complete there always will be a $V$ operator if $H$ has an antilinear symmetry \cite{footnote2}, with $\langle \psi(t)\vert V\vert \psi(t)\rangle$ being the most general inner product that one could introduce that is probability conserving. If we introduce right-eigenvectors $\vert R\rangle$ of $H$ according to  $H\vert R\rangle=E\vert R \rangle$, we have
\begin{align}
\langle R\vert H^{\dagger}=E^*\langle R\vert=\langle R\vert VHV^{-1},\qquad \langle R\vert VH=E^*\langle R\vert V.
\label{2.5}
\end{align}
Consequently, we  can identify a left eigenvector $\langle L\vert=\langle R\vert V$, and can write the inner product as $\langle R\vert V\vert R\rangle=\langle L\vert R\rangle$. Thus in general we can  identify the left-right inner product as the most general probability-conserving inner product in the antilinear case, a form that could perhaps be  anticipated since a Hamiltonian cannot have any more eigenvectors than its left and right ones. Thus the way to generalize Hermitian quantum mechanics is to replace the dual space that is built out of the Hermitian conjugate bras $\langle R\vert$ of a given set of $\vert R\rangle$ kets by a dual space built on  $\langle R\vert V=\langle L\vert$ bras instead. Moreover, while this antilinear requirement recovers standard probability-conserving Hermitian quantum mechanics when $V=I$, for $V\neq I$ it is not just a more general probability-conserving requirement, it is actually the most general \cite{Mannheim2018a}, and thus has to be used when  the Hamiltonian is not Hermitian. 

\section{Application of $PT$ Symmetry}
\label{S3}

In the discussion given in Sec. \ref{S1}  we considered scattering by a well and obtained a time delay, with there being only one pole, viz. that at $E_0-i\Gamma$, while our discussion of $PT$-symmetric Hamiltonians that we provide in this section will lead us to poles at both  $E_0-i\Gamma$ and  $E_0+i\Gamma$.  We now show that the two approaches lead to the same imaginary part for the associated propagators. To this end we follow \cite{Mannheim2013,Mannheim2018a} and consider a typical  non-Hermitian $2\times 2$ matrix (a simple $C$-conserving analog of a two-energy-level atomic system)
\begin{align}
M(s)=\begin{pmatrix}
1+i&s\\ 
s&1-i
\end{pmatrix},
\label{3.1}
\end{align}
where the parameter $s$ is real and positive. The matrix $M(s)$ does not obey the Hermiticity condition $M_{ij}=M^*_{ji}$. However, if we set $P=\sigma_1$ and  $T=K$, where $K$ denotes complex conjugation we obtain $PTM(s)[PT]^{-1}=M(s)$, with $M(s)$ thus being  $PT$ symmetric for any value of  the real parameter $s$. With the eigenvalues of $M(s)$ being given by $E_{\pm}=1 \pm (s^2-1)^{1/2}$, we see that both of these eigenvalues are real if $s$ is either greater or equal to one, and form a complex conjugate pair if $s$ is less than one, just as is to be expected with $PT$ symmetry. 

For the complex conjugate eigenvalue case with $s<1$ we can diagonalize $M(s)$ via $S=\cosh\alpha+\sigma_2 \sinh\alpha$ with $\tanh(2\alpha)=s$ \cite{footnote3}, so as to give
\begin{align}
M^{\prime}(s<1)&=SM(s<1)S^{-1}=\begin{pmatrix}1+i(1-s^2)^{1/2}&0\\ 0&1-i(1-s^2)^{1/2}\end{pmatrix}
=\begin{pmatrix}E_0+i\Gamma&0\\ 0&E_0-i\Gamma \end{pmatrix}.
\label{3.9}
\end{align}
The diagonal $M^{\prime}(s<1)$ given in (\ref{3.9}) is $PT$ symmetric under $P=\sigma_1$, $T=K$. There will thus be a relevant $V$, and it is given by $V=-i\sigma_2$. With this $V$,  and with $E_0=1$, the eigenkets, eigenbras and  orthogonality and closure relations are given by 
\begin{align}
&u_+=e^{-it+\Gamma t}\begin{pmatrix}1\\ 0\end{pmatrix}, \qquad u_-=e^{-it-\Gamma t}\begin{pmatrix}0\\ 1\end{pmatrix}, \qquad u_+^{\dagger}V=e^{it+\Gamma t}\begin{pmatrix}0&-1 \end{pmatrix},\qquad u^{\dagger}_-V=e^{it-\Gamma t}\begin{pmatrix}1 &0\end{pmatrix},
\nonumber\\
&u_{\pm}^{\dagger}Vu_{\pm}=0,\qquad u_{-}^{\dagger}Vu_{+}=+ 1,\qquad u_{+}^{\dagger}Vu_{-}=- 1,
\qquad u_{+}u^{\dagger}_{-}V-u_{-}u^{\dagger}_{+}V=I.
\label{3.3}
\end{align}
The appearance of the $-1$ factor in $u_{+}^{\dagger}Vu_{-}$ is not indicative of any possible negative norm ghost problem since  $u_{+}^{\dagger}Vu_{-}$ is a transition matrix element between two different states, and not the overlap of a state with its own conjugate.
As we see, all of the $V$-based inner products are time independent, with the only nonvanishing ones being the ones that link the decaying and growing modes. As the population of one level decreases the population of the other level increases by the same amount, with probability conservation requiring that we consider both levels together and not just one (the decaying one) as is usually done in the  standard Breit-Wigner approach. 
Given (\ref{3.3}) the associated propagator is  given by
\begin{align}
D_{PT}(E)=\frac{u_{-}^{\dagger}Vu_{+}}{E-(E_0-i\Gamma)}+\frac{u_{+}^{\dagger}Vu_{-}}{E-(E_0+i\Gamma)}=\frac{1}{E-(E_0-i\Gamma)}-\frac{1}{E-(E_0+i\Gamma)}.
\label{3.4}
\end{align}
Thus we obtain
\begin{align}
D_{PT}(E)=\frac{-2i\Gamma}{(E-E_0)^2+\Gamma^2},
\label{3.5}
\end{align}
to thus give the same negative imaginary sign for the propagator as obtained in the standard (\ref{1.2}) that only contained one pole, so that operationally (\ref{3.5}) and (\ref{1.2})  are equivalent. It is this equivalence that  enabled Lee and Wick \cite{Lee1969} to consider the relativistic generalization of (\ref{3.4}), viz. a propagator of the form $1/(k^2-M^2+iN^2)-1/(k^2-M^2-iN^2)$, in order to obtain better large $k^2$ behavior (viz. $1/k^4$) than a standard $1/k^2$ propagator, to  then control quantum field theory renormalization \cite{footnote4}.

While $D_{PT}(E)$ gives the energy dependence of the propagator,  how we determine its time dependence, viz. 
\begin{align}
D_{PT}(t)=\frac{1}{2\pi }\int^{\infty}_{-\infty} dEe^{-iEt}D_{PT}(E),
\label{3.6}
\end{align}
depends on the form of the  contour that we choose in the complex $E$ plane. There are two poles in $D_{PT}(E)$, one above the real $E$ axis and one  below, and we can suppress the lower-half circle contribution when $t>0$, and suppress the upper-half circle contribution when $t<0$. However, in order to be able to combine the two propagators in (\ref{3.4}) so as to give the propagator given in (\ref{3.5}), as noted in \cite{Lee1969}  we need the two poles in $D_{PT}(E)$ to be in the same contour. We thus deform the contour around the upper-half plane pole at $E=E_0+i\Gamma$ so that it contributes when we close below. And with both poles then being in the same contour, we obtain
\begin{align}
D_{PT}(t)=-i\theta(t)[e^{-iE_0t-\Gamma t}-e^{-iE_0t+\Gamma t}].
\label{3.7}
\end{align}
Thus we obtain both a forward in time growing mode and a forward in time decaying mode. Thus we obtain a time-advanced, negative-width generalization of the positive-width time delay produced by a potential well. We can associate the time advance  with an energy-dependent phase shift of the form $\tan \delta=-\Gamma/(E_0-E)$, a phase shift equal to $-\pi/2$ at resonance, together with a scattering amplitude and time advance of  the form  
\begin{align}
f(E)\sim e^{i\delta}\sin\delta=- \frac{\Gamma}{E_0+i\Gamma-E},\qquad \Delta T(E)=\hbar\frac{d\delta}{dE}=-\frac{\hbar\Gamma}{[(E-E_0)^2+\Gamma^2]},\qquad \Delta T(E_0)=-\frac{\hbar}{\Gamma}.
\label{3.8}
\end{align}
Thus in $PT$-symmetric quantum mechanics  a time advance (viz. a negative time delay) is natural.

Rather than work with matrices we can also write a wave equation with solutions $e^{-iE_0t+\Gamma t}$ and $e^{-iE_0t-\Gamma t}$ with opposite signs for $\Gamma$. Specifically, setting $E=id/dt$ in $(E-E_0-i\Gamma)(E-E_0+i\Gamma)=0$ we obtain
\begin{align}
\left[\frac{d^2}{dt^2}+2iE_0\frac{d}{dt}-E_0^2-\Gamma^2\right]\psi(t)=0.
\label{3.12}
\end{align}
In contrast, for  a damped harmonic oscillator with energy $E=E_0-i\Gamma$ and  wave function $e^{-iE_0t-\Gamma t}$ we have
\begin{align}
\left[\frac{d^2}{dt^2}+2\Gamma\frac{d}{dt}+E_0^2+\Gamma^2\right]\psi(t)=0.
\label{3.13}
\end{align}
Since (\ref{3.13}) is a second-order differential equation it has a second solution of the form $e^{iE_0t-\Gamma t}$ with the opposite sign for $E_0$, a solution that is also damped rather than growing. In energy space (\ref{3.13}) is written as $(E-E_0+i\Gamma)(E+E_0+i\Gamma)=0$.  However, the $PT$ symmetry wave equation associated with (\ref{3.12}) provides for both time delay and time advance rather than just the standard Hermitian theory time delay \cite{footnote7}. 

In order to be able to apply this time advance to the excitation and then decay of an atom, we must identify the two eigenvalues in (\ref{3.9}) not as those of the ground state and the excited state  but as the two transition energies, one to excite and populate the upper level and the other to decay and depopulate the upper level, with both transitions being caused by the virtual photon clouds in the natural line widths of the atomic energy levels.  With $PT$ symmetry the associated time delay and time advance will cancel each other identically. In the many  $PT$-symmetry experiments that have been performed the emphasis has been on systems with  gain and loss that is balanced, i.e., $E_0+i\Gamma$ and $E_0-i\Gamma$ with the same $\Gamma$. Balancing time advance and time delay is thus in the same vein \cite{footnote6}. Within  $\hbar/E_0\approx 10^{-15}$ seconds the emission of the photon by the upper level  in the atomic case thus essentially occurs at the same  instant as the initial absorption of a photon by the lower level with no $10^{-9}$  seconds time delay \cite{footnote8}, 
\cite{footnote9}. In the experiments of \cite{Sinclair2022,Angulo2024}, which use ultracold Rubidium atoms,  it would appear that an effect of this type  has been detected, with \cite{Angulo2024} also reporting a measured negative time delay.

\begin{acknowledgments}
The author wishes to acknowledge helpful discussions with C. M. Bender, A. R. P. Rau, S. Rotter and J. Sinclair, and wishes to thank P. Guo for informing him of his and his collaborators' recent work on negative time delay \cite{Guo2022,Guo2023}.
\end{acknowledgments}


\begin{thebibliography}{99}
\bibitem{footnote1} In the optics literature it is conventional to discuss a quantity  $-d\delta/dE$, which is known as the group delay.  Far from resonance both it and the time delay are positive. Near a resonance they have opposite sign, so a negative group delay at resonance still means a positive time delay, not a negative one.
\bibitem{footnote6} The $PT$-symmetry time advance discussed here is not the same as the time advance phenomenon  that  occurs in Hermitian quantum mechanics. In this latter case the center of a wave packet approaching  a well (or barrier) travels on the classical trajectory. However, the wings of the wave packet not only spread out they can start to deform due the presence of the well before the center of the wave packet actually reaches the well.
\bibitem{footnote7} In addition, we note that in the limit $\Gamma \rightarrow 0$ there is also a sharp difference between the solutions to (\ref{3.12}) and (\ref{3.13}). The solutions to the non-$PT$-symmetric (\ref{3.13}) are the two stationary wave functions $e^{-iE_0t}$ and $e^{iE_0t}$ with nondegenerate energies $E_0$ and $-E_0$. In contrast,  for the $PT$-symmetric (\ref{3.12}) while both energies become the same $E_0$, the two wave functions collapse onto just the one $e^{-iE_0t}$. The second solution is given by the $\Gamma \rightarrow 0$  limit of $[e^{-iE_0t+\Gamma t }-e^{-iE_0t-\Gamma t}]/2\Gamma$, to thus be of the nonstationary form $te^{-iE_0t}$. With only one stationary eigenfunction rather than two the $PT$-symmetric case becomes of nondiagonalizable Jordan-block form.
\bibitem{footnote8} With the causal $\theta(t)$ factor in (\ref{3.7}) an excited state does not emit a photon before the ground state absorbs it, with the time advance not being bigger than the time delay.
\end{thebibliography}
\end{document}